\documentclass[letter]{ptptex}

\usepackage{graphicx}

\title{$1/N$ Expansion of the 2D $CP^{N-1}$ Model on Non(anti)commutative Superspace}

\author{Kazutoshi \textsc{Araki},$^{1,}$\footnote{Address before April 1.} Takeo \textsc{Inami},$^{1,}$\footnote{E-mail: inami@phys.chuo-u.ac.jp} Hiroaki \textsc{Nakajima}$^{2,}$\footnote{E-mail: nakajima@skku.edu} \\ and Yasuhiro \textsc{Yamashita}$^{1,\ast)}$}

\inst{$^{1}$Department of Physics, Chuo Universty, Tokyo 112-8851, Japan \\ $^{2}$Department of Physics and Institute of Basic Science, Sungkyunkwan University, Suwon 440-746, Korea}

\abst{We study UV properties of the two-dimensional supersymmetric $CP^{N-1}$ model on non(anti)commutative superspace. We show that the deformed model has the same UV property as the ordinary model in the leading order of $1/N$ and the deformation give rise to UV divergence which cannot be dealt with in the next-to-leading order.}

\begin{document}

\maketitle

\noindent
\textit{1. Introduction}
\hspace*{24pt}
Recently theories defined on non(anti)commutative(NC) superspace have been studied extensively~\cite{Seiberg:2003yz, d-susy}, due to their realization in the string theory by turning on a constant graviphoton field strength $F^{\alpha \beta}$, which changes the anticommutation relation between fermionic variables of the superspace. In four-dimensions(4D), the non(anti)commutativity is introduced to $\mathcal{N}=1$ superspace by deforming the anticommuting relations for superspace coordinates $\theta^{\alpha}$ so that they may satisfy Clifford algebra:
\begin{equation}
	\{ \theta^{\alpha},\,\theta^{\beta} \}=C^{\alpha \beta} \label{eq:anticom}.
\end{equation}
$C^{\alpha \beta}$ is NC parameter, and this deformation is called $C$-deformation. The remarkable characteristics of field theories on NC superspace are that half of $\mathcal{N}=1$ supersymmetry(SUSY) is broken. The left, unbroken symmetry is called $\mathcal{N}=1/2$ SUSY. 

Supersymmetric field theories have a softer ultraviolet(UV) behavior than those without SUSY. UV divergences become softer in theories with extended SUSY($\mathcal{N} \ge 2$) than in theories with $\mathcal{N}=1$ SUSY. It is interesting to investigate whether and how UV properties of SUSY field theories will change by deforming $\mathcal{N}=1$ SUSY to $\mathcal{N}=1/2$. The renormalizability question has been studied for the $\mathcal{N}=1$ super Yang-Mills theory~\cite{NCYM} and the Wess-Zumino model~\cite{NCWZ} in perturbation and by a more general argument~\cite{Britto:2003aj}.

In this letter we study the renormalizability question of the 2D $\mathcal{N}=2$ SUSY $CP^{N-1}$ (nonlinear) sigma model on NC superspace. The same question has been studied low order of perturbation~\cite{Araki:2006nn}, implying the same UV property as the non-deformed SUSY model. Here we study the question in the $1/N$ expansion. In this method propagation of auxiliaty fields will change by $C$-deformation and they may yield different UV divergence properties. We will carry out the computation to the next-to-leading order.

\noindent
\textit{2. The Leading Order}
\hspace*{24pt}
The 2D SUSY $CP^{N-1}$ model on NC superspace can be obtained from the 4D one~\cite{Inami:2004sq} by dimensional reduction. The Lagrangian is given by
\begin{align}
	\mathcal{L}=
	  &-\bar{\phi_{l}}\,D^{2} \phi_{l}+(\sigma^{2}+\pi^{2})\,\bar{\phi_{l}}\,\phi_{l}-\alpha \left( \bar{\phi_{l}}\,
	    \phi_{l}-\frac{N}{g} \right)+i \bar{\psi_{l}} {D \hspace*{-7pt} /}\,\psi_{l} \notag \\
	  &\hspace*{36pt} +\bar{\psi_{l}} \left( \sigma+i \gamma^{5} \pi \right) \psi_{l}+\bar{c}\,(\psi_{l}\,\bar{\phi_{l}})
	    +\left(\phi_{l}\,\bar{\psi_{l}}-i \psi_{l}^{\mathrm{T}} C {\partial \hspace*{-6pt} /} \bar{\phi_{l}} \right) c\,.
	    \label{eq:lagrangian}
\end{align}
The model contains of $N$ complex scalar fields $\phi_l$ and $N$ Dirac fields $\psi_l\,(l=1,\,2,\,\ldots,\,N)$, which we refer to as the matter fields. In addition we have auxiliary fields $A_{\mu},\,\alpha,\,\sigma,\,\pi,\,c$. Susy field theories on NC superspace are formulated in Euclidean spacetime. In this approach $\bar{\phi}_l$ is not the Hermitian conjugate of $\phi_l$ and is an independent field. The same remark applies to $\bar{\psi}_l$ and $\bar{c}$.

The generating functional of the model is
\begin{align}
	Z=\int \mathcal{D} \phi_l\,\mathcal{D} \bar{\phi_l}\,\mathcal{D} \psi_l\,\mathcal{D} \bar{\psi_l}\,
	  &\mathcal{D} \sigma\,\mathcal{D} \pi\,\mathcal{D} \alpha\,\mathcal{D} c\,\mathcal{D} A_{\mu} \notag \\
	  &\times \exp \left\{ -\int d^{2} \hspace*{-2pt} x\,\left( \mathcal{L}+\bar{J_l}\,\phi_l+\bar{\phi_l}\,J_l+\bar{\eta_l}\,
	    \psi_l+\bar{\psi_l}\,\eta_l \right) \right\}.
\end{align}
Performing the integrations over the matter fields $\phi_l,\,\bar{\phi}_l,\,\psi_l$ and $\bar{\psi}_l\,$, we obtain
\begin{equation}
	Z=\int \mathcal{D} \sigma\,\mathcal{D} \pi\,\mathcal{D} \alpha\,\mathcal{D} c\,\mathcal{D} \bar{c}\,\mathcal{D} A_{\mu}
	  \exp (-NS_{\mathrm{eff}}).
\end{equation}
The effective action for the auxiliary fields $S_{\mathrm{eff}}$ is given by
\begin{equation}
	S_{\mathrm{eff}}=\mathrm{Tr}\,\ln \left[ \Delta_{\phi}-\bar{c}\,\Delta_{\psi}^{-1} c-ic\,C {\partial \hspace*{-6pt} /}\,
	    \Delta_{\psi}^{-1} c \right]-\mathrm{Tr}\,\ln \Delta_{\psi}+\frac{1}{g} \int d^{2} \hspace*{-2pt} x\,\alpha\,+\mbox{(source)},
	    \label{eq:eaction}
\end{equation}
where $\Delta_{\phi} \equiv -D^{2}+\sigma^{2}+\pi^{2}-\alpha^{2}\,,\Delta_{\psi} \equiv i({D \hspace*{-7pt} /}+\gamma^{5} \pi)+\sigma$.

Performing the Legendre transform of $S_{\mathrm{eff}}$ and setting all fields to constants, we obtain the effective potential :
\begin{align}
	V_{\mathrm{eff}}=
	  &\,\bar{v} v\,\left( \langle \alpha \rangle-\langle \sigma \rangle^{2}-\langle \pi \rangle^{2} \right)
	    -\frac{\langle \alpha \rangle}{g} \notag \\
	  &\qquad -\int \frac{d^{2} k}{(2 \pi)^{2}}\,\ln \left( k^{2}-\langle \sigma \rangle^{2}+\langle \pi
	    \rangle^{2}-\langle \alpha \rangle \right) \notag \\
	  &\qquad +\int \frac{d^{2} k}{(2 \pi)^{2}}\,\mathrm{Tr}\,\ln \left( {k \hspace*{-5.0pt} /}+i \gamma^{5}
	    \langle \pi \rangle+\langle \sigma \rangle \right)\,,
\end{align}
where the vacuum expectation value(VEV) of $\phi_l$ and $\bar{\phi}_l$ are written as $\langle \phi \rangle=(0,\,0,\,\ldots,\,\frac{v}{\sqrt{N}})$, $\langle \bar{\phi} \rangle=(0,\,0,\,\ldots,\,\frac{\bar{v}}{\sqrt{N}})$. The VEVs of all fields other than the scalar fields are set to zero. The vacuum of this model is determined by the stationary conditions for this potential :
\begin{align}
	\frac{\partial V_{\mathrm{eff}}}{\partial v}
	  &=\bar{v}\,\left( \langle \alpha \rangle-\langle \sigma \rangle^{2}-\langle \pi \rangle^{2} \right)=0\,, \\
	\frac{\partial V_{\mathrm{eff}}}{\partial \bar{v}}
	  &=v\,\left( \langle \alpha \rangle-\langle \sigma \rangle^{2}-\langle \pi \rangle^{2} \right)=0\,, \\
	\frac{\partial V_{\mathrm{eff}}}{\partial \langle \sigma \rangle}
	  &=-2 \bar{v} v\,\langle \sigma \rangle-\int \frac{d^{2} k}{(2 \pi)^{2}}\,\frac{2 \langle \sigma \rangle}{k^{2}
	    +\langle \sigma \rangle^{2}+\langle \pi \rangle^{2}-\langle \alpha \rangle} \notag \\
	  &\hspace*{84pt} +\int \frac{d^{2} k}{(2 \pi)^{2}}\,\frac{2 \langle \sigma \rangle}{k^{2}
	    +\left( i \gamma^{5} \langle \pi \rangle+\langle \sigma \rangle \right)^{2}}=0 \label{eq:cond1}\,, \\
	\frac{\partial V_{\mathrm{eff}}}{\partial \langle \pi \rangle}
	  &=-2 \bar{v} v\,\langle \pi \rangle-\int \frac{d^{2} k}{(2 \pi)^{2}}\,\frac{2 \langle \pi \rangle}{k^{2}
	    +\langle \sigma \rangle^{2}+\langle \pi \rangle^{2}-\langle \alpha \rangle} \notag \\
	  &\hspace*{84pt} -\int \frac{d^{2} k}{(2 \pi)^{2}}\,\frac{2 \langle \pi \rangle}{k^{2}+\left( i \gamma^{5}
	    \langle \pi \rangle+\langle \sigma \rangle \right)^{2}}=0 \label{eq:cond2}\,, \\
	\frac{\partial V_{\mathrm{eff}}}{\partial \langle \alpha \rangle}
	  &=\bar{v} v\,-\frac{1}{g}+\int \frac{d^{2} k}{(2 \pi)^{2}}\,\frac{1}{k^{2}+\langle \sigma \rangle^{2}+\langle \pi
	    \rangle^{2}-\langle \alpha \rangle}=0\,.
\end{align}
We look for the supersymmetric vacuum and set $\langle \pi \rangle=0$, $\langle \alpha \rangle=0$. Then, the eq. (\ref{eq:cond2}) is fulfilled automatically, and the other conditions are
\begin{gather}
	\bar{v}\,\langle \sigma \rangle^{2}=0\,, \label{eq:cond3} \\
	v\,\langle \sigma \rangle^{2}=0\,, \label{eq:cond4} \\
	2\,\bar{v} v\,\langle \sigma \rangle=0\,, \label{eq:cond5} \\
	\bar{v} v-\frac{1}{g}+\int \frac{d^{2} k}{(2 \pi)^{2}}\,\frac{1}{k^{2}+\langle \sigma \rangle^{2}}=0\,. \label{eq:cond6}
\end{gather}
The solution to eq. (\ref{eq:cond3}), (\ref{eq:cond4}) and (\ref{eq:cond5}) are
\begin{equation}
	v=\bar{v}=0 \quad \text{or} \quad \langle \sigma \rangle=0\,.
\end{equation}
$\langle \sigma \rangle=0$ corresponds the $SU(N)$ broken phase. But in 2D field theories, the broken phase is forbidden by Coleman's theorem~\cite{}. Hence there is only the $SU(N)$ symmetric phase ($v=\bar{v}=0, \langle \sigma \rangle \neq 0$).

In the symmetric phase, the eq. (\ref{eq:cond6}) is
\begin{equation}
	\bar{v} v-\frac{1}{g}+\frac{1}{2 \pi} \ln \left( \frac{\Lambda}{\langle \sigma \rangle} \right)=0\,, \label{eq:cond6_2}
\end{equation}
where $\Lambda$ is the momentum cut-off. This divergence is dealt with by renormalizing the coupling constant as
\begin{equation}
	\frac{1}{g}=\frac{1}{g_{\mathrm{R}}}+\frac{1}{2 \pi} \ln \left( \frac{\Lambda}{\mu} \right)\,,
\end{equation}
where $g_{\mathrm{R}}$ is the renormalized coupling constant, $\mu$ is the renomalization scale. Then, the eq. (\ref{eq:cond6_2}) reduces to
\begin{equation}
	\bar{v} v-\frac{1}{g_{\mathrm{R}}}+\frac{1}{2 \pi} \ln \left( \frac{\mu}{\langle \sigma \rangle} \right)=0\,,
	  \label{eq:cond6_3}
\end{equation}
and we get the VEV of $\sigma$
\begin{equation}
	\langle \sigma \rangle=\mu \exp \left( -\frac{2 \pi}{g_{\mathrm{R}}} \right) \equiv m(\mu,\,g_{\mathrm{R}})\,.
\end{equation}

The $\beta$-function is obtained from eq. (\ref{eq:cond6_3}),
\begin{equation}
	\beta (g_{\mathrm{R}}) \equiv \mu \frac{\partial g_{\mathrm{R}}}{\partial \mu}=-\frac{g_{\mathrm{R}}^{2}}{2 \pi}\,.
\end{equation}
This result is the same as the ordinary 2D SUSY $CP^{N-1}$ model. We have found that the deformed model has the same UV property as the ordinary model in the leading order of $1/N$.

Propagators of auxiliary fields are required in computing diagrams in the next-to-leading order. The model contains four kinds of auxiliary fields : $\alpha$, $\sigma$, $\pi$, $c$ and $A_{\mu}$. They all begin to propagate after taking into account the quantum effects of $\phi_l$ and $\psi_l$ loops. Considering the fact that $\sigma$ acquires a non-zero VEV($\langle \sigma \rangle =m$), we perform the shift
\begin{equation}
	\sigma \rightarrow \sigma+m,\quad \langle \sigma \rangle=0\,. \label{eq:shift1}
\end{equation}
The fields $\alpha$ and $\sigma$ mix as they propagate. Hence, it is convenient to diagonalize their propagators by rewriting $\alpha$ as
\begin{equation}
	\alpha \rightarrow \alpha +2m \sigma\,. \label{eq:shift2}
\end{equation}
Performing the shift (\ref{eq:shift1}) and (\ref{eq:shift2}) in the Lagrangian (\ref{eq:lagrangian}), we obtain
\begin{align}
	\mathcal{L}=
	  & \,\bar{\phi_{l}}\,(-D^{2}+m^{2}) \phi_{l}+(\sigma^{2}+\pi^{2})\,\bar{\phi_{l}}\,\phi_{l}-\alpha
	    \left( \bar{\phi_{l}}\,\phi_{l}-\frac{N}{g} \right) \notag \\
	  & \hspace*{36pt} +\frac{N}{g} \cdot 2m \sigma+i \bar{\psi_{l}} ({D \hspace*{-7pt} /}+m)\,\psi_{l}
	    +\bar{\psi_{l}} \left( \sigma+i \gamma^{5} \pi \right) \psi_{l}\\
	  & \hspace*{36pt} +\bar{c}\,(\psi_{l}\,\bar{\phi_{l}})+\left( \phi_{l}\,\bar{\psi_{l}}-i \psi_{l}^{\mathrm{T}} C
	    {\partial \hspace*{-6pt} /} \bar{\phi_{l}} \right) c\,. \notag
\end{align}
The propagators of $\phi_l$ and $\psi_l$ are the same as the ordinary model :
\begin{equation}
	D_{ij}^{\phi}=\frac{\delta_{ij}}{p^{2}+m^{2}}\,,\quad D_{ij}^{\psi}=\frac{\delta_{ij}}{-{p \hspace*{-4pt} /}+m}\,.
\end{equation}
The effective propagators of $\alpha$, $\sigma$, $\pi$, $c$ and $A_{\mu}$ can be obtained from the effective action after redefining the fields $\sigma$ and $\alpha$ as (\ref{eq:shift1}) and (\ref{eq:shift2}). They all have the factor $1/N$ and are given by
\begin{gather}
	D^{\alpha}=-\frac{1}{NI(p)}\,,\quad D^{\sigma}=\frac{1}{N} \frac{1}{p^{2} I(p)}\,,\quad
	  D^{\pi}=-\frac{1}{N} \frac{1}{(p^{2}+8m^{2})I(p)-\frac{1}{\pi}}\,, \notag \\
	D^{\bar{c} c}=-\frac{1}{N} \frac{2}{({p \hspace*{-4pt} /}-2m)\,I(p)}\,,\quad
	  D^{cc}=-\frac{1}{N} \frac{1}{m{p \hspace*{-4pt} /}C\,I(p)}\,, \notag \\
	D^{A}_{\mu \nu}=\frac{1}{N} \frac{1}{p^{2}I(p)} \left\{ \delta_{\mu \nu}-(1-\xi) \frac{p_{\mu} p_{\nu}}{p^{2}} \right\}
\end{gather}
where $I(p)$ is defined by
\begin{equation}
	I(p) \equiv \frac{1}{2 \pi m^{2} (z^{2}-1)^{1/2}} \ln z\,,\quad z \equiv \frac{p^{2}+2m^{2}}{2m^{2}}
	 \,,
\end{equation}
and $\xi$ is the $U(1)$ gauge parameter. $D^{cc}$ appears due to $C$-deformation and does not exist in the ordinary model. This $D^{cc}$ propagator can replace $D^{\bar{c} c}$ in all the diagrams that exist in the ordinary model, giving rise to new diagrams without due to $C$-deformation. We study the effects of these new diagrams.

\noindent
\textit{3. The Next-to-Leading Order}
\hspace*{24pt}
Our purpose is to study the (non)renormalizability of the deformed SUSY $CP^{N-1}$ model. The question is whether any of new graphs give rise to UV divergence which cannot be absorbed by terms that are already contained in the Lagrangian.

\begin{figure}[htbp]
	\begin{center}
		\includegraphics[width=10cm,height=13cm,clip]{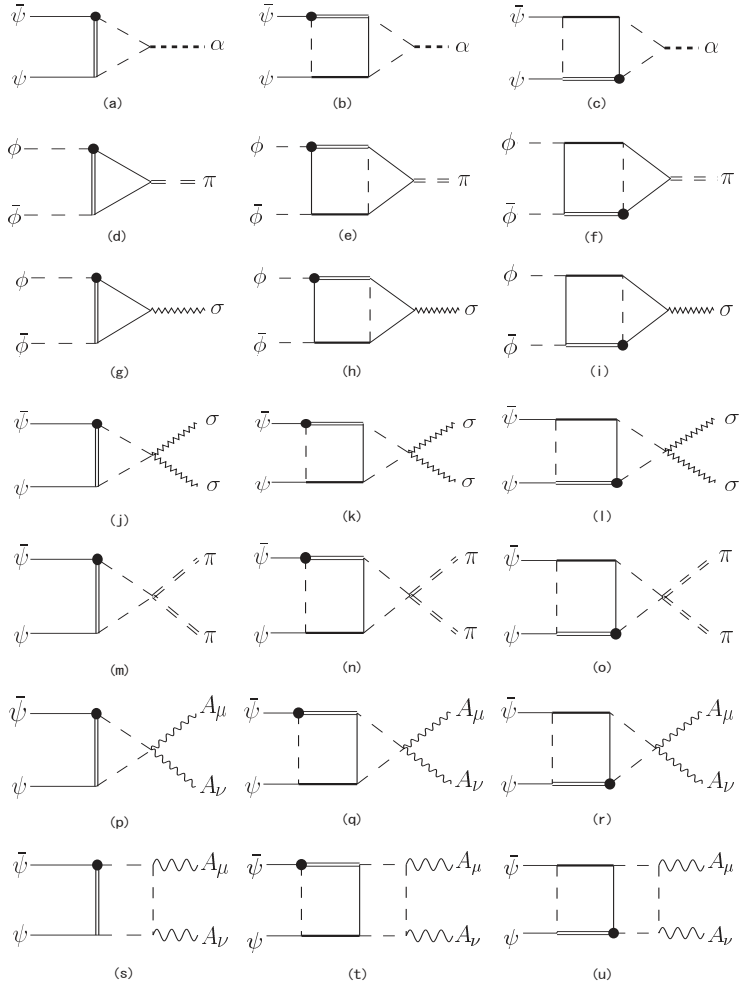}
		\caption{one- and two-loops graphs containing $D^{cc}$}
		\label{fig:diagram}
	\end{center}
\end{figure}

We studied graphs which contain the new propagator $D^{cc}$. There are twenty-one graphs at one and two loops(Fig.\ref{fig:diagram}). The dotted line, the thick line and the double line express $D^{\alpha}, D^{\bar{c} c}$ and $D^{cc}$, respectively. Graphs of higher loops turn out to be beyond the next-to-leading order. Of the twenty-one graphs only six contain UV divergence : (a), (g), (i), (j), (m), (p). After loop momentum integration, their UV divergence behaviors are
\begin{gather}
	\Gamma_{\alpha \bar{\psi} \psi} \sim \frac{1}{4mN} \ln \left| \frac{\ln \tilde{\Lambda}}{\ln \tilde{\mu}} \right|\,,\quad
	  \Gamma_{\pi \bar{\phi} \phi}=0,\\
	\Gamma_{\sigma \bar{\psi} \psi} \sim -\frac{m}{N} \left[ 17 \left( \,\mathrm{Li}(\tilde{\Lambda})-\mathrm{Li}(\tilde{\mu})\,
	  \right)-\frac{9}{2} \ln \left| \frac{\ln \tilde{\Lambda}}{\ln \tilde{\mu}} \right|+3 \left(
	  \frac{\tilde{\Lambda}}{\ln \tilde{\Lambda}}-\frac{\tilde{\mu}}{\ln \tilde{\mu}} \right) \right]\,,\\
	\Gamma_{\sigma^{2} \bar{\psi} \psi},\,\Gamma_{\pi^{2} \bar{\psi} \psi} \sim
	  \frac{1}{2mN} \ln \left| \frac{\ln \tilde{\Lambda}}{\ln \tilde{\mu}} \right|\,,\quad \Gamma_{A^{2} \bar{\psi} \psi}=0,\label{eq:divergence_C2}
\end{gather}
where $\tilde{\Lambda},\,\tilde{\mu}$ are defined by $\tilde{\Lambda} \equiv (\Lambda^{2}+2m^{2})/2m^{2},\,\tilde{\mu} \equiv (\mu^{2}+2m^{2})/2m^{2}$. Li(x) is logarithmic integral. To summarize, four vertex functions diverge: $\alpha \bar{\psi} \psi$, $\sigma \bar{\phi} \phi$, $\sigma^{2} \bar{\psi} \psi$, $\pi^{2} \bar{\psi} \psi$. These terms are not contained in the Lagrangian.

\noindent
\textit{4. Summary}
\hspace*{24pt}
We have studied the ultraviolet properties of the 2D SUSY $CP^{N-1}$ model on NC superspace using the 1/$N$ expansion. In the leading order, we have calculated the $\beta$-function of this model and found that 2D SUSY $CP^{N-1}$ model on NC superspace has the same UV property as the ordinary model. We have also found that the new propagator, $D^{cc}$ appears due to $C$-deformation.

In the next-to-leading order of 1/$N$, new diagrams appears by existence of $D^{cc}$. We studied graphs which contain the new propagator $D^{cc}$ and found that four vertex functions diverge: $\alpha \bar{\psi} \psi$, $\sigma \bar{\phi} \phi$, $\sigma^{2} \bar{\psi} \psi$, $\pi^{2} \bar{\psi} \psi$. The Lagrangian cannot absorb these UV divergences. In order to deal with these divergences, we have to add these terms to the Lagrangian as counter terms by hands. These counter terms affect the UV properties in beyond the next-to-leading order. It is necessary to carry out the computation to the higher order.

\section*{Acknowledgements}
This paper is supported partially by research grants of Japanese
Ministry of education and science, tokutei (18034006),
kiban A (18204024), kiban B (16340040), kiban C.
This paper is also
supported by
the Postdoctoral Research Program of Sungkyunkwan University (2007)
and is the result of research activities (Astrophysical Research
Center for the Structure and Evolution of the Cosmos (ARCSEC))
supported by KOSEF (H.~N.).

%

\bigskip

\end{document}